\RequirePackage{fix-cm}
\documentclass[smallextended]{svjour3}       
\smartqed  
\usepackage{graphicx}
\usepackage{mathptmx}      
\usepackage{latexsym}
\usepackage{bm}
\usepackage{subcaption}
\usepackage{hyperref}
\usepackage[mathscr]{euscript}
\usepackage{color}
\begin{document}

\title{A path-integral formulation of the run and tumble motion and chemotaxis in {\it Escherichia coli}
}

\titlerunning{Path-integral formulation of run and tumble motion}        

\author{C. S. Renadheer        \and
        Ushasi Roy \and 
        Manoj Gopalakrishnan
}


\institute{\at 
              Department of Physics, IIT Madras,\\
              Chennai 600036, India.\\
              Tel.: +91-44-2257 4894\\
              Fax: +91-44-2257 4852\\
              \email{manojgopal@iitm.ac.in}           
}

\date{Received: date / Accepted: date}

\maketitle

\begin{abstract}
Bacteria such as {\it Escherichia coli} move about in a series of runs and tumbles: while a run state (straight motion) entails all the 
flagellar motors spinning in counterclockwise mode,  a tumble is caused by a shift in the state of one or more motors to clockwise 
spinning mode. In the presence of an attractant gradient in the environment, runs in the favourable direction are extended, and this 
results in a net drift of the organism in the direction of the gradient. The underlying signal transduction mechanism produces directed 
motion through a bi-lobed response function which relates the clockwise bias of the flagellar motor to temporal changes in the attractant 
concentration. The two lobes (positive and negative) 
of the response function are separated by a time interval of $\sim 1$s, such that the bacterium effectively compares the concentration at 
two different positions in space and responds accordingly. We present here a novel path-integral method which allows us to address this 
problem in the most general way possible, including multi-step CW-CCW transitions, directional persistence and power-law waiting time 
distributions. The method allows us to calculate quantities such as the effective diffusion coefficient and drift velocity,  
in a power series expansion in the attractant gradient. Explicit results in the 
lowest order in the expansion are presented for specific models, which, wherever applicable, agree with the known results. New results for 
gamma-distributed run interval distributions are also presented. 
\end{abstract}

\vspace{2cm}
\section{Introduction}
\label{intro}

The run and tumble motion of the bacterium {\it Escherichia coli} ({\it E. coli}) is well-known, and is characterised by a series of straight runs, interspersed 
with shorter tumbles, during which the organism reorients itself. In the absence of a chemoattractant (and when the concentration of 
chemoattractant is uniform), the motion is globally unbiased in space. When an attractant gradient is present, the bacterium extends 
its runs in the favourable direction. As a result, a net drift develops in the direction of the gradient, which enables the organism 
to move towards the source. The resulting chemotaxis in bacterium is, therefore,  fundamentally different from similar phenomena in 
unicellular eukaryotes like amoeba and neutrophils, in which moving cells reorient themselves in the favourable direction by sensing 
concentration difference across their body\cite{bibHCBerg}. For a review of the experimental literature and a summary of the modelling 
approaches, we refer the reader to \cite{bibMaini}. 

In the simplest kinematic description of the run and tumble motion, it is assumed that, in the absence of 
chemoattractants, tumbling is a first order process characterised by a single rate $R$ specifying 
the switch from counter-clockwise (CCW) to clockwise (CW) rotation of a single flagellar motor, which initiates the tumbling process. 
In a static organism, it is observed that a stimulus in the form of a time-dependent change in attractant concentration $\delta c(t)$ 
introduces a corresponding change in the clockwise bias, which may be expressed mathematically through a linear response relation of the form

\begin{equation}
\delta R(t)=-R\int_{0}^{t}\chi(t-t^{\prime})\delta c(t^{\prime})dt^{\prime}
\label{eqA}
\end{equation}

where $\chi(t)$ is a linear response function. Experiments\cite{bibHCBerg} have shown that at least for some attractants, the area enclosed by the 
response function is near-zero\cite{bibSegallBerg}, which implies that the organism adapts perfectly to a step-like increase of stimulus. 
In this case, the response function has a two-lobe structure, with a positive lobe appearing almost immediately after the stimulus is applied, 
and a negative lobe appearing later, with the centres of the lobes being separated by nearly a second. These properties led de Gennes 
\cite{bibdeGennes} to suggest the approximate form 

\begin{equation}
\chi(t)\simeq \kappa\left[\delta(t)-\delta(t-\Delta)\right]
\label{eqB}
\end{equation} 
where $\Delta$ is the time delay between the centres of the positive and negative lobes and $\kappa$ is an empirical parameter which depends 
on the details of the underlying biochemical network. The response function has also been computed explicitly \cite{bibReneaux,bibURoy} 
using variants of the Barkai-Leibler model\cite{bibBarkai} for the receptor methylation-demethylation processes, originally introduced to 
explain the perfect adaptation property of {\it E. coli}, and the robustness of the network output to cell-to-cell variations in enzyme 
concentrations. 

Using a combination of heuristic arguments and rigorous calculations, de Gennes \cite{bibdeGennes} derived the following expression for the drift velocity 
of a bacterium in two dimensions, in a concentration gradient $\nabla c=\boldsymbol{\alpha}$:

\begin{equation}
{\bf v}_d\sim D{\boldsymbol \alpha}\int_{0}^{\infty}\chi(t)e^{-Rt}dt
\label{eqC}
\end{equation}
Here $D=v^2/2R$ is the diffusion coefficient for the unbiased run and tumble walk, $v$ being the run speed. Later, other authors have 
also derived expressions identical to Eq.\ref{eqC}. Melissa and Gopalakrishnan \cite{bibReneaux} followed an approach similar to 
de Gennes \cite{bibdeGennes}  (discussed in detail in the next section) to compute the drift velocity, but derived the response function 
directly from a simplified version of the Barkai-Liebler model\cite{bibBarkai} for receptor methylation and demethylation. Celani and Vergassola \cite{bibCelani} obtained the response function from a general Fokker-Planck equation for the run and tumble motion, with a finite number of abstract internal variables included, thus  ensuring that the function has the required bilobe form. The latter paper also generalised the result in  \cite{bibdeGennes} 
to arbitrary spatial dimensions and also included directional persistence between successive runs. 

The motivations behind this paper are as follows. The original derivation of Eq.\ref{eqC} by de Gennes \cite{bibdeGennes} includes a 
simplifying step, which although justifiable {\it a posteriori}, can be avoided, we feel. Specifically, instead of computing the 
mean displacement due to a gradient as a single average over possible trajectories, de Gennes computes first the mean displacement due to
the gradient over a single run event, and multiplies it with the mean number of runs over a certain time interval to find the asymptotic 
drift velocity. As a result, the generality of the result, i.e., Eq.\ref{eqC}, is not apparent, which has also led to suggestions that 
the result applies only under the condition $\Delta\ll R^{-1}$ (although it was also noted that numerical simulations indicate otherwise) 
\cite{bibSakuntala}. It is also generally assumed that the run intervals are exponentially distributed and are typically longer than tumbles.  However,  experimental observations\cite{bibKorobkovaPRL,bibParkBiophysJ} indicate that time  intervals corresponding to clockwise and counterclockwise modes of rotation of a single motor are, in general, gamma-distributed, the details of which depend on the mean clockwise bias of the motor. It has also been suggested that tumbles correspond to rigid body (Brownian) rotations of the bacterium in space \cite{bibSaragosti}.  

The contents of this paper, therefore, were borne out of our attempts to develop a mathematically rigorous formulation of the run and tumble 
motion and chemotaxis in bacteria such as {\it E.coli}, which could potentially include non-exponentially distributed run (and tumble) intervals, directional correlations between successive runs and  finite tumble intervals. The path-integral formalism for run and tumble motion and chemotaxis in bacteria presented here offers the following advantages. An individual trajectory is specified using the set of run and tumble intervals, and angles specifying the directions of runs. The probability distributions of run and tumble intervals are specified using two cumulative (survival) probability functions, while 
directional persistence is introduced through a conditional probability density connecting the direction of the present run with that of 
the previous one. A probability functional for a trajectory is constructed using all these quantities, using which any desired statistical 
average can be computed systematically. The formalism works well as, and has the structure of, a systematic perturbation theory in which the imposed attractant gradient $\alpha$ is the small parameter. We show that, as special cases of interest, many standard results can be reproduced using our 
approach, including (a) Eq.\ref{eqC} , (b) super and sub-diffusive motion when run intervals are power-law distributed and (c) modification of 
the effective diffusion coefficient by correlation between directions of successive runs. We also present our new results for 
bacterial drift velocity when run intervals are gamma-distributed, as indicated by recent experimental data \cite{bibParkBiophysJ}. 

A note on the historical origins of the formalism: different variants of the technique have been used earlier,  in the study of ion channel dynamics 
\cite{bibGoychuk,bibGopalakrishnan}, reaction-diffusion processes in cells \cite{bibBenichou} as well as search and capture of chromosomes 
by microtubules \cite{Gopalakrishnan:2011aa}. Here, we develop and expand it further to study the run and tumble motion in bacteria. Although only results obtained to the lowest (non-trivial) order in $\alpha$ are presented here, we stress that, in principle, the computation could be extended to any higher power of the same. Furthermore, all the calculations are presented here for spatial dimension $d=2$; however, generalisations to $d=1$ and $d=3$ are straightforward.

\section{The general path-integral functional for run and tumble motion}
\label{sec:1}


Let us consider a bacterium executing run and tumble motion in two dimensions. Denote by $f(T)$, the cumulative probability of run intervals while 
$g(\tau)$ shall be the same for tumbles, so that $f(0)=g(0)=1$ by definition. The probability distributions of the run and tumble interval durations, respectively, are $-\dot{f}$ and $-\dot{g}$. 

A time interval $[0:t]$ could fall into one of the two following categories: (a) $N$ completed runs, $N$ completed tumbles, 
and one incomplete run ($N\geq 0$), or (b) $N$ completed runs, $N-1$ completed tumbles, and one incomplete tumble ($N\geq 1$). 
Let $T_i$ denote the time intervals corresponding to runs, and $\tau_i$ denote the same for tumbles. During a tumble, the bacterium 
undergoes reorientation such that its direction of motion changes. Let $G_{\tau}(\theta|\theta_0)$ denote the probability distribution 
of the final angle $\theta$, given initial angle $\theta_0$ and time of tumble $\tau$. Let $\theta_i$ be the angle specifying the run with duration $T_i$. We choose the convention
that the bacterium always starts in a run state at $t=0$, and the duration of the first run is $T_1$, while that of the first tumble is $\tau_1$. Then, the probability functionals 
describing paths corresponding to situation (a) and (b) are, respectively, 

\begin{equation}
\Phi_N^{(a)}({\bf T},{\bm\tau},{\bm\theta};t)=f(T_{N+1})\delta\left(T_{N+1}+\sum_{i=1}^{N}(T_i+\tau_i)-t\right)\psi(\theta_1)
\prod_{i=1}^{N}{\dot f}(T_i){\dot g}(\tau_i)G_{\tau_i}(\theta_{i+1}|\theta_i)
\label{eq1a}
\end{equation}
and 
\begin{equation}
\Phi_N^{(b)}({\bf T},{\bm\tau},{\bm\theta};t)=-g(\tau_N){\dot f}(T_N)\delta\left(\sum_{i=1}^{N}(T_i+\tau_i)-t\right)
\psi(\theta_1)\prod_{i=1}^{N-1}{\dot f}(T_i){\dot g}(\tau_i)G_{\tau_i}(\theta_{i+1}|\theta_i),
\label{eq1b}
\end{equation}
where $\psi(\theta_1)=(2\pi)^{-1}$ is the probability distribution of the initial angle $\theta_1$. The functionals are normalized as follows:

\begin{equation}
\sum_{N=0}^{\infty}\int_a \Phi_N^{(a)}({\bf T},{\bm\tau},{\bm\theta};t){\mathcal D}T {\mathcal D}{\tau} {\mathcal D}{\theta}
+\sum_{N=1}^{\infty}\int_b \Phi_N^{(b)}({\bf T},{\bm\tau},{\bm\theta};t){\mathcal D}T{\mathcal D}{\tau}{\mathcal D}{\theta}=1,
\label{eq2}
\end{equation}
where 
\begin{eqnarray}
\int_a...{\mathcal D}T{\mathcal D}{\tau} {\mathcal D}{\theta}\equiv \int_{0}^{T}...dT_1\int_{0}^{2\pi}d\theta_1\int_{0}^{T-T_1}d\tau_1\int_{0}^{T-T_1-\tau_1}dT_2\int_{0}^{2\pi}d\theta_2.....\times\nonumber\\
\int_{0}^{t-\sum_{i=1}^{N}T_i-\sum_{j=1}^{N-1}\tau_j}d\tau_{N}\int_{0}^{t-\sum_{i=1}^{N}T_i-\sum_{j=1}^{N}\tau_j}dT_{N+1}\int_{0}^{2\pi}d\theta_{N+1}
\nonumber
\end{eqnarray}
and 
\begin{eqnarray}
\int_b...{\mathcal D}T {\mathcal D}{\tau} {\mathcal D}{\theta}\equiv \int_{0}^{T}...dT_1\int_{0}^{2\pi}d\theta_1\int_{0}^{T-T_1}d\tau_1\int_{0}^{T-T_1-\tau_1}dT_2\int_{0}^{2\pi}d\theta_2.....\times\nonumber\\
\int_{0}^{t-\sum_{i=1}^{N-1}T_i-\sum_{j=1}^{N-1}\tau_j}dT_{N}\int_{0}^{2\pi}d\theta_N\int_{0}^{t-\sum_{i=1}^{N}T_i-\sum_{j=1}^{N-1}\tau_j}d\tau_{N}
\nonumber
\end{eqnarray}
are time-ordered integrals. Using the above functionals, the mean of any dynamical quantity which depends explicitly on the 
variables $\{{\bf T}, {\bm \tau},{\bm\theta}\}$ may be calculated. Let $a(t)$ be such a quantity (e.g., the net displacement), 
whose value for a given trajectory may be denoted ${\mathcal A}_{N}({\bf T}, {\bm \tau}, {\bm \theta})$\cite{bibvankampen}. Then, the ensemble average of $a$ is given by 

\begin{equation}
\overline{a}(t)= \langle {\mathcal A}_N({\bf T}, {\bm \tau},{\bm \theta})\rangle=  \langle {\mathcal A}_N({\bf T}, {\bm \tau},{\bm \theta})\rangle_a+ \langle {\mathcal A}_N({\bf T}, {\bm \tau}, {\bm \theta})\rangle_b,
\label{eq2a}
\end{equation}
where 

\begin{eqnarray}
\langle {\mathcal A}_N({\bf T}, {\bm \tau},{\bm \theta})\rangle_a=\sum_{N=0}^{\infty}\int_a  {\mathcal A}_{N}({\bf T},{\bm\tau},{\bm\theta})\Phi_N^{(a)}({\bf T},{\bm\tau},{\bm\theta};t){\mathcal D}T{\mathcal D}{\tau} {\mathcal D}{\theta}\nonumber\\
\langle {\mathcal A}_N({\bf T}, {\bm \tau},{\bm \theta})\rangle_b=\sum_{N=1}^{\infty}\int_b {\mathcal A}_{N}({\bf T},{\bm\tau},{\bm\theta})\Phi_N^{(b)}({\bf T},{\bm\tau},{\bm\theta};t){\mathcal D}T {\mathcal D}{\tau}{\mathcal D}{\theta}
\label{eq3a}
\end{eqnarray}

Furthermore, the probability distribution for the variable $a$ may be expressed as 

\begin{eqnarray}
P(a,t)=\langle \delta(a-{\mathcal A}_{N}({\bf T},{\bm\tau},{\bm\theta}))\rangle.
\label{eq3b}
\end{eqnarray}

In the following subsection, we will discuss a simplified version of the general model introduced here, which will be used for mathematical calculations. 


\subsection{The ``minimal'' model: Instantaneous tumbles, no directional correlations}

In this simplified model, which shall serve as a ``reference'' model for us, (i) the tumble durations are assumed to be negligibly small in comparison with run durations ( i.e., $\tau_i\to 0$ everywhere), and (ii) directional correlations between successive runs are ignored. Condition (i) requires that we choose the tumble time distribution to be $-{\dot g}(\tau)=\delta(\tau)$. This implies that $g(\tau)=0$ for $\tau\neq 0$ while $g(0)=1$. It is then 
clear that in this case, events corresponding to class (b) in Eq.\ref{eq2a} and Eq.\ref{eq3a} do not contribute in the averaging, and may be ignored. Condition (ii) implies that $G_{\tau_i}(\theta_{i+1}|\theta_i)=(2\pi)^{-1} \forall i\in[1,N]$ in Eq. \ref{eq1a}. As a result, the probability functional in Eq.\ref{eq1a} reduces to (the superscript ``m'' denoting ``minimal'' from now on) 

\begin{equation} 
\Phi_{N}^{(m)}({\bf T},{\bm\theta};t)=\frac{1}{(2\pi)^{N+1}}f(T_{N+1})\delta\bigg(T_{N+1}+\sum_{i=1}^{N}T_i-t\bigg)\prod_{i=1}^{N}(-1)^N{\dot f}(T_i).
\label{eq5}
\end{equation}

We will first show explicitly that the functional in Eq.\ref{eq5} is normalised as in Eq.\ref{eq2}. Define formally the ``normalisation integral'' 
${\cal N}(t)=\sum_{N=0}^{\infty}\int \Phi_{N}^{(m)}({\bf T},{\bm\theta};t) {\mathcal D}T{\mathcal D}{\theta}$. To evaluate the r.h.s,  we use Eq.\ref{eq4} (after putting  ${\mathcal A}\equiv 1$), to find 

\begin{eqnarray}
{\cal N}(t)\equiv \sum_{N=0}^{\infty}\int \Phi^{(m)}_N({\bf T},{\bm\theta};t) {\mathcal D}T{\mathcal D}{\theta}=\frac{1}{(2\pi)^{N+1}}\sum_{N=0}^{\infty} \int d\omega_1 ...d\omega_{N+1}\nonumber \\
\prod_{i=1}^N i\omega_i F(\omega_i)e^{-i\sum_{i=1}^N\omega_i T_i} 
F(\omega_{N+1}) e^{-i\omega_{N+1}(t-\sum_{i=1}^N T_i)}
\label{eq5+}
\end{eqnarray}

where $F(\omega)$ is the Fourier transform of $f(T)$, as defined in the previous section. Next, we Laplace-transform Eq.\ref{eq5+}, and use the convolution theorem mentioned in Sect.~\ref{sec:1} to find



\begin{equation} 
{\mathcal L}_{s}[{\cal N}]=\frac{1}{(2\pi)^{N+1}}\int  \frac{d\omega F(\omega)}
  {s+i\omega} \sum_{N=0}^{\infty} \Bigg[ \int  \frac{d\omega i\omega F(\omega)}{s+i\omega} \Bigg]^N
\label{eq5++}
\end{equation}

After completing the elementary geometric sum, and noting that $\int_{-\infty}^{\infty} d\omega F(\omega)\equiv f(0)=1$ by definition, 
it follows that ${\mathcal L}_{s}[{\cal N}]=s^{-1} $, and hence ${\cal N}(t)=1$ as required.


The average of any dynamical quantity $a\equiv {\mathcal A}_N({\bf T},{\bm \theta})$ associated with the 
motion may be evaluated as ${\overline a}(t)=\langle  {\mathcal A}_N({\bf T},{\bm \theta})\rangle_m$, where 

\begin{equation}
\langle  {\mathcal A}_N({\bf T},{\bm \theta})\rangle_m=\sum_{N=0}^{\infty}\int {\mathcal D}T{\mathcal D}\theta \Phi_{N}^{(m)}({\bf T},{\bm\theta};t) {\mathcal A}_N({\bf T},{\bm \theta})
\label{eq5+++}
\end{equation}

When $f(T)$ is non-exponential,  it is convenient to express Eq.\ref{eq5+++} using its Fourier transform $F(\omega)$, defined by the relation $2\pi F(\omega)=\int_{-\infty}^{\infty}f(T)e^{i\omega T}dT$ (with the understanding that $f(T)=0$ for $T<0$). The result is 

\begin{eqnarray}
\langle {\mathcal A}_N({\bf T},{\bm \theta})\rangle_m=\sum_{N=0}^{\infty}\frac{1}{(2\pi)^{N}}\int_a  {\mathcal A}_{N}({\bf T},{\bm\theta}){\mathcal D}T{\mathcal D}{\theta}\int_a {\mathcal D}{\omega}e^{-i\omega_{N+1}t}F(\omega_{N+1})\times\nonumber\\ 
\prod_{i=1}^{N}i\omega_i F(\omega_i)e^{-i(\omega_i-\omega_{N+1})T_i}
\label{eq4}
\end{eqnarray}

where $\int...{\mathcal D}{\omega}\equiv \int ...\prod_{i=1}^{N+1}d\omega_i$. 

\subsection{Special cases of the minimal model}

In this subsection, we introduce a few special cases of the functional in Eq.\ref{eq5}, as well as a few extensions. 

\paragraph{(a) Exponentially distributed run intervals (exponential model)}:

A particularly simple special case of the minimal model is that of exponentially distributed run durations, 
where $-{\dot f}(T)=Re^{-RT}H(T)$  and $-{\dot g}(\tau)=\delta(\tau)$, where $H(T)$ is the Heaviside step-function: $H(T)=$1 for $T\geq 0$ and 0 otherwise . This implies $f(T)=e^{-RT}H(T)$ and the corresponding probability functional is 

\begin{equation}
\Phi_{N}^{(e)}({\bf T},{\bm\theta};t)=\frac{1}{(2\pi)^{N+1}}e^{-R T_{N+1}}\delta\left(T_{N+1}+\sum_{i=1}^{N}T_i-t\right)R^N e^{-R\sum_{i=1}^{N} T_i}
\label{eq6a}
\end{equation}

In the above expression, the superscript ``e'' denotes ``exponential''. 

\paragraph{(b) Power-law distributed run intervals - L\'evy flights}:

Earlier experimental observations by Korobkova et al.\cite{bibKorobkova} had suggested that the cumulative probability of CCW interval durations of a single flagellar motor shows power-law decay for nearly two decades in time. Partly motivated by this observation, we study a model with $f(T)=(1+\gamma T)^{-\beta}$ with $\gamma>0$ and $\beta>0$. 

\subsection{Extensions of the minimal model}

\paragraph{(c) Exponential model with directional correlations between successive runs}:

The run and tumble motion of {\it E. coli} observed in experiments is characterized by directional persistence, i.e., the directions 
of two consecutive runs are positively correlated. For the sake of simplicity, we assume here that the correlation exists only between 
two consecutive runs. The probability functional for this case is a straightforward generalisation of Eq.\ref{eq6a}: 

\begin{equation}
\Phi_{N}^{(e,p)}({\bf T},{\bm\theta};t)=e^{-R T_{N+1}}\delta\left(T_{N+1}+\sum_{i=1}^{N}T_i-t\right)R^N e^{-R\sum_{i=1}^{N} T_i} \psi(\theta_1)\prod_{j=1}^{N} G(\theta_{j+1}|\theta_j),
\label{eq5b1}
\end{equation}

where the additional superscript ``p'' represents persistence/antipersistence in run directions, and $\psi(\theta_1)=(2\pi)^{-1}$ as mentioned earlier (since the initial run direction is chosen randomly). To bring in directional correlations between successive runs, we choose 

\begin{equation}
G(\theta_{j+1}|\theta_j)=\frac{1}{2\pi}\bigg( 1+J\cos(\theta_{j+1}-\theta_j)\bigg)~~~~ \forall j \in [1,N]
\label{eq5b2}
\end{equation}

In the above expression, $J$ is a phenomenological parameter to be chosen such that $|J|<1$ to ensure 
positivity of $G(\theta_{j+1}|\theta_j)$. Further, $J > 0 $ implies persistence and $J < 0$ implies anti-persistence of motion. 
It is also easily verified that $\langle \cos(\theta_{j+1}-\theta_j)\rangle=J/2$, so that the parameter $J$ may be fixed using the value of the average in the l.h.s, as observed 
in experiments. 

\paragraph{(d) Exponential model with chemotaxis}:  

In {\it E. coli}, chemotaxis is achieved by making the tumble rate a function of the previous positions of the bacterium. For a general 
path and time-dependent tumble rate $R(t)\equiv R(t;{\bf T},{\bm\theta})$, Eq.\ref{eq5} may be generalised as 

\begin{equation}
\Phi_{N}^{(e,c)}({\bf T}, {\bm\theta};t)=\frac{1}{(2\pi)^{N+1}}e^{-\int_{t_{N}}^{t}R(T)dT}\prod_{i=1}^{N}R(t_{i})e^{-\int_{t_{i}}^{t_{i+1}}R(T)dT}
\label{eq7}
\end{equation}

where $t_i=\sum_{j=1}^{i}T_j$ for $1\leq i\leq N$. In the minimal model with exponentially distributed run intervals, the dependence of 
tumbling rate on attractant concentration may be expressed through the linear response relation

\begin{equation}
R(T)=R\left[1-\int_{0}^{T}\chi(T-T^{\prime})c[\vec{r}(T^{\prime})]dT^{\prime}\right]
\label{eq8}
\end{equation}

which follows directly from Eq.\ref{eqA}. Here, $R$ is the tumble rate in the absence of attractant and ${\bf r}(t)$ is the position 
of the bacterium at time $t$. In the case of a uniform attractant gradient such that $\nabla c({\bf r})={\bm \alpha}$, 
we have $c({\bf r})={\bm\alpha}\cdot {\bf r}$. Without loss of generality, we choose ${\bm\alpha}=\alpha{\hat {\bm x}}$, such that 

\begin{equation}
R(T)=R\left[1-\alpha\int_{0}^{T}\chi(T-T^{\prime})x(T^{\prime})dT^{\prime}\right]
\label{eq8a}
\end{equation}

Eq.\ref{eqB}, when substituted in Eq.\ref{eq8a}, leads to the following ``path-dependent'' tumble rate: 

\begin{equation}
R(T)=R\left[1-\kappa\alpha(x(T)-x(T-\Delta))\right]
\label{eq10}
\end{equation}

Eq.\ref{eq10}, when used in Eq.\ref{eq7} leads to the following expansion of the probability functional, in the limit of weak gradient:

\begin{eqnarray}
\Phi_{N}^{(e,c)}({\bf T}, {\bm\theta};t)=\frac{1}{(2\pi)^{N+1}}R^N e^{-Rt}\bigg[1+\alpha\bigg(\kappa R\int_{t-\Delta}^{t}x(T)dT-\nonumber\\
\kappa\sum_{j=1}^{N}[x(t_j)-x(t_j-\Delta)]+{\mathcal O}(\alpha)\bigg)\bigg],
\label{eq11}
\end{eqnarray}

where the second superscript, ``c'' indicates ``chemotaxis''. In the following section, we present our important results for each of these models. The calculations use the following ``theorem'' extensively, which is a straightforward generalisation of the standard convolution theorem in Laplace transforms. 

\noindent
{\it Theorem:} Given a function $f(t)=\int_{0}^{t}dT_1h_1(T_1)\int_{0}^{t-T_1}dT_2h_2(T_2).....\int_{0}^{t-\sum_{i=1}^{N-1}T_i}dT_N h_N(T_N)$, 
its Laplace transform is ${\mathcal L}_s[f]=s^{-1}\prod_{i=1}^{N}{\mathcal L}_s[h_i]$. For the (rather elementary) proof, we refer the 
reader to \cite{bibGopalakrishnan}.

\section{Results}
\label{sec:2}

\subsection{Mean-square displacement in the minimal model: general results}

The mean square displacement (MSD) in the minimal model is given by $\langle r^2\rangle=\langle {\bf R}_N({\bf T},{\bm \theta})\cdot {\bf R}_N({\bf T},{\bm \theta})\rangle_m$, where 

\begin{equation}
\mathcal {\bf R}_N({\bf T},{\bm \theta})=v\sum_{i=1}^{N+1}T_i {\bf e}_i
\label{eqRN}
\end{equation}

is the displacement vector for a certain path, with ${\bf e}_i= {\bf i}\cos\theta_i + {\bf j} \sin\theta_i$ being unit vectors specifying directions of individual runs. After using Eq.\ref{eq5+++}, the following general expression for the Laplace transform of the MSD is arrived at, after a straightforward computation using the 
convolution theorem: 

\begin{equation}
{\mathcal L}_s[\langle r^2\rangle]=\frac{2v^2}{s^2}\frac{I_2(s;F)}{I_1(s;F)} 
\label{eq5a}
\end{equation}

where the integrals are defined as 

\begin{equation} \label{I1I3}
I_1(s;F)=\int \frac{d\omega F(\omega)}{s+i\omega}~~;~~ I_2(s;F)=\int \frac{d\omega i\omega F(\omega)}{(s+i\omega)^3}.
\label{eq5b}
\end{equation}

(It is also useful to note that $I_2(s;F)=I_3^{\prime\prime}(s;F)/2$, where $I_3(s;F)=1-sI_1(s;F)$ and the prime denotes differentiation with respect to $s$). As a special case, it follows that, if $\lim_{s\to 0+} I_2/I_1$ is 
non-zero and finite, $\langle r^2\rangle\sim 4Dt$, with the diffusion coefficient $D$ being given by the general expression

\begin{equation}
D=\frac{v^2}{2} \lim_{s\to 0+}\frac{I_2(s;F)}{I_1(s;F)}
\label{eq6}
\end{equation}

\subsection{Mean-square displacement for specific models}

\paragraph{(a) Exponential model}:\\

The diffusion coefficient can be easily found using the expression in Eq.\ref{eq6}. Here, $2\pi F(\omega)=(R-i\omega)^{-1}$, which leads to the diffusion coefficient 

\begin{equation}
D_e=v^2/2R
\label{eq6+}
\end{equation}

for the exponential model. In Sect.~\ref{sec:5a}, we also show  explicitly that the probability distribution of the displacement in the long-time, large distance limit, for this model, is Gaussian. 

\paragraph{(b) Power-law distributed run intervals}: \\

The details of the calculations are presented in Sect.\ref{sec:5c}. A summary of the results are given below. 

\begin{eqnarray}
\langle r^2\rangle &\propto& t^2 ~~~~~~~~~ 0< \beta \leq 1 \nonumber\\
\langle r^2\rangle &\propto& t^{3-\beta} ~~~~ 1< \beta \leq 2 \nonumber\\
\langle r^2\rangle &\propto& t ~~~~~~~~~~ \beta >  2
\label{eq6b}
\end{eqnarray}

Thus, the motion is ballistic for $\beta\leq 1$,  super-diffusive when $1< \beta<2$ and diffusive when $\beta\geq 2$. The above results agree 
with the predictions made in \cite{bibBouchad}, derived using heuristic scaling arguments ( a slightly different model is presented in \cite{bibThiel}, where both run and tumble intervals are assumed to be power-law distributed). 

\paragraph{(c) Model with directional correlations between runs}:\\

After carrying out the required calculations (see Sect.~\ref{sec:5b} for details) we find $\langle r^2\rangle\sim 4D_J t$, where 

\begin{equation}
D_J=\frac{v^2}{2R(1-J/2)}
\label{eq5b11}
\end{equation}

is the diffusion coefficient for the run and tumble walk, when directional persistence is present. The expression in 
Eq.\ref{eq5b11} agrees with the more general expression of Celani and Vergassola \cite{bibCelani}, derived by a different method for arbitrary 
spatial dimension $d$. 

\subsection{Chemotaxis in {\it E. coli}: mean displacement and drift velocity}

We now use the functional in Eq.\ref{eq11} to compute the mean displacement of the bacterium in the long-time limit, 
and thereby derive an expression for the drift velocity to  the lowest order in $\alpha$. It is easily seen that, 
in the evaluation of $\langle x(t)\rangle$, the leading term (${\mathcal O}(\alpha^0)$) does not contribute in the long-time limit, 
and the leading non-zero term can be written as the sum of two terms: $\langle x(t)\rangle=x_1(t)+x_2(t)$ with 

\begin{eqnarray}
x_{1}(t) &=& \alpha\kappa R\int_{t-\Delta}^{t}\langle x(t^{\prime})x(t)\rangle_{e} dt^{\prime}\label{eq12a}\\
x_{2}(t) &=& -\alpha\kappa\sum_{j=1}^{N}[\langle x(t)x(t_j)\rangle_m-\langle x(t)x(t_j-\Delta)\rangle_e]
\label{eq12}
\end{eqnarray}

where the averages need to be carried out using the functional in Eq.\ref{eq6a}. The first average is particularly simple; this is 
because for the unbiased run and tumble walk, we expect $\langle x(t)x(t^{\prime})\rangle\sim 2D_et^{\prime}$ for $t^{\prime}\leq t$ 
in the large $t$-limit, similar to Brownian diffusion\cite{bibvankampen}, with the diffusion coefficient $D_e$ being given by Eq.\ref{eq6+}. 
Substituting this result in Eq.\ref{eq12a} leads to the asymptotic result $x_1(t)\sim v_1t$, with 

\begin{equation}
v_1=\alpha\kappa v^2 \Delta 
\label{eq12+}
\end{equation}

The computation of $x_2(t)$ is more involved, and the details are to be found in the following subsection. 

\subsection{Calculation of $x_2(t)$ in Eq.\ref{eq12}}
\label{sec:3d}


%
%

Let us start with Eq. \ref{eq12}, and express the r.h.s in the form

\begin{equation}
x(t_j)-x(t_j-\Delta)=\sum_{i=1}^{j}P_i^{j}\cos\theta_i,
\label{eq5d1}
\end{equation}
where 

\begin{eqnarray}
P_{i}^{j}=v\bigg(\Delta-\sum_{q=i+1}^{j}T_q\bigg)\bigg[ H \bigg( \Delta-\sum_{q=i+1}^{j}T_q\bigg)-H \bigg( \Delta-\sum_{m=i}^{j}T_m\bigg)\bigg]
\nonumber\\
+vT_iH\bigg(\Delta-\sum_{m=i}^{j}T_m\bigg);\quad\quad 1\leq i\leq j-1
\label{eq5d2}
\end{eqnarray} 

while 

\begin{equation}
P_{i}^{i}=v\Delta H(T_i-\Delta)+vT_iH(\Delta-T_i), 
\label{eq5d3}
\end{equation}

where the Heaviside step-function $H(t)$ has been defined earlier, before Eq.\ref{eq6a}. Next, we define the integrals


\begin{equation}
I_m=-v\alpha \kappa \int \frac{1}{(2\pi)^{N+1}} \mathcal{D}\theta\int \mathcal{D}T\bigg(\sum_{j=1}^{N}[x(t_j)-x(t_j-\Delta)]\bigg) T_m\cos\theta_m
\label{eq5d4}
\end{equation}
such that 
\begin{equation}
x_2(t)= \sum_{N=0}^{\infty}R^Ne^{-Rt} \bigg[ \sum_{m=1}^{N+1}I_m \bigg].
\label{eq5d5}
\end{equation}
The angular integrations in Eq.\ref{eq5d4} are easily done using Eq. \ref{eq5d1}-\ref{eq5d3}. The result is 

%

\begin{equation}
I_m =\sum_{r=m}^{N}I_m^r,
\label{eq5d6}
\end{equation}
where 

\begin{equation}
I_m^r=-\frac{ v \alpha \kappa}{2}\int \mathcal{D}T P_{m}^{r}T_m
\label{eq5d7}
\end{equation}
The time ordered integral in Eq.\ref{eq5d7} can be done using the theorem in Sect.~\ref{sec:1}, after expressing the Heaviside 
functions in Eq.\ref{eq5d2} using the integral representation $2\pi H(\Delta-t)=\int_{0}^{\Delta}dy\int_{-\infty}^{\infty}d 
\psi e^{i\psi(y-t)}$. Thus, it follows that 

\begin{eqnarray}
\mathcal{L}_{s}[I_m^r]=\frac{1}{2\pi}\frac{\Delta}{s^{N+m-r+2}}\int_{0}^{\Delta}dy\int_{-\infty}^{\infty}e^{i\psi y} d \psi (s+i\psi)^{m-r}\
\nonumber\\
-\frac{1}{2\pi}\frac{(r-m)}{s^{N+m-r+2}}\int_{0}^{\Delta}dy\int_{-\infty}^{\infty}e^{i\psi y} d \psi (s+i\psi)^{m-r-1}
\nonumber\\
-\frac{1}{2\pi}\frac{\Delta}{s^{N+m-r}}\int_{0}^{\Delta}dy\int_{-\infty}^{\infty}e^{i\psi y} d \psi (s+i\psi)^{m-r-2}\
\nonumber\\
+\frac{1}{2\pi}\frac{(r-m+2)}{s^{N+m-r}}\int_{0}^{\Delta}dy\int_{-\infty}^{\infty}e^{i\psi y} d \psi (s+i\psi)^{m-r-3}
\label{eq5d8}
\end{eqnarray}
The integrals in the above expression are straightforward; for any $n\geq 0$, we have 
$\frac{1}{2\pi}\int_{0}^{\Delta}dy\int_{-\infty}^{\infty}d \psi e^{i\psi y} (s+i\psi)^{-n} 
=\int_{0}^{\Delta}y^{n-1}e^{-sy}dy$, hence 


%
%

\begin{equation}
\mathcal{L}_{s}[I_{N-h}]=\sum_{j=1}^{h+1}\beta_j(s),
\label{eq5d9}
\end{equation}
where, after the rescaling $sy\equiv \phi$, 

%
%
%
%

\begin{eqnarray}
\beta_j(s)=\frac{\Delta}{s^{N+2}}\int_{0}^{\Delta s} e^{-\phi}\phi^{j-2}d\phi-\frac{(j-1)}{s^{N+3}}\int_{0}^{\Delta s} e^{-\phi}\phi^{j-1}d\phi
\nonumber\\
-\frac{\Delta}{s^{N+2}}\int_{0}^{\Delta s} e^{-\phi}\phi^{j}d\phi+\frac{(j+1)}{s^{N+3}}\int_{0}^{\Delta s} e^{-\phi}\phi^{j+1}d\phi~~~~2\leq j\leq N
\label{eq5d10}
\end{eqnarray}
while 

\begin{equation}
\beta_1(s)=\frac{\Delta}{s^{N+2}}-\frac{\Delta}{s^{N+2}}\int_{0}^{\Delta s} e^{-\phi}\phi d\phi+\frac{2}{s^{N+3}}\int_{0}^{\Delta s} e^{-\phi}\phi^2d\phi
\label{eq5d11}
\end{equation}
Using Eq.\ref{eq5d9} in Eq.\ref{eq5d5}, it follows that 

\begin{equation}
\mathcal{L}_s[ x_2(t)]= \sum_{N=0}^{\infty}R^N \bigg[ \sum_{j=1}^{N}(N-j+1)\beta_j(s+R) \bigg]
\label{eq5d12}
\end{equation}
The sum in the r.h.s of the above equation, after some algebra, is found to be 

\begin{equation}
\sum_{j=1}^{N}(N-j+1)\beta_j(s)=\frac{N\Delta}{s^{N+2}}(2-e^{-\Delta s})-\frac{N}{s^{N+3}}\int_{0}^{\Delta s}e^{-\phi}\phi d\phi.
\label{eq5d13}
\end{equation}

After substituting Eq.\ref{eq5d13} into Eq.\ref{eq5d12} and completing the sum, we find that, in the limit $s\to 0$, 

\begin{equation}
\mathcal{L}_s[ x_2(t)] \sim \frac{\kappa v^2 \alpha}{2R s^2}(1 -e^{-\Delta R}-2\Delta R)
\label{eq5d14}
\end{equation}

which directly leads to the asymptotic result $x_2(t)\sim v_2 t$, where 

\begin{equation}
v_2=\frac{\kappa v^2\alpha}{2R}\left(1-e^{-\Delta R}-2\Delta R\right). 
\label{eq13}
\end{equation}

Adding Eq.\ref{eq12+} and Eq.\ref{eq13} leads to the complete result $\langle x(t)\rangle\sim v_d t$, where the drift velocity $v_d=v_1+v_2$ is given by 

\begin{equation} \label{xt}
v_d=\frac{\kappa v^2\alpha}{2R}\left(1-e^{-\Delta R}\right)
\label{eq14}
\end{equation}

in agreement with de Gennes \cite{bibdeGennes}, and is a special case of the more general expression in Eq.\ref{eqC}, when the response function is 
approximated as in Eq.\ref{eqB}. This follows from the following argument. To derive Eq.\ref{eqC} from Eq.\ref{eq14}, note that, according to Eq. \ref{eq14}, 
a response function $\chi(t)=\delta(t-\Delta)$ 
would result in a drift ``velocity'' $v_d=D_e\alpha e^{-R\Delta}$. Now, we may express 
$\delta(t-\Delta)=(2\pi)^{-1}\int_{-\infty}^{\infty}e^{i\omega(t-\Delta)}d\omega$. Since any arbitrary response function may be expressed as 
$\chi(t)=\int_{-\infty}^{\infty} {\tilde \chi}(\omega)e^{i\omega t}d\omega$, it follows that the general expression for 
drift velocity should be $v_d=2\pi D_e\alpha{\tilde \chi}(-iR)$, in agreement with Eq.\ref{eqC} (note that $\chi(t)=0$ for $t<0$). 

In Sect.~\ref{sec:5d}, we show how the result in Eq.\ref{eq14} is generalised for arbitrarily distributed run intervals.

\subsection{An application: Drift velocity and diffusion coefficient for gamma-distributed run durations}

Recent experimental observations\cite{bibKorobkovaPRL,bibParkBiophysJ} have shown that durations of CW and CCW intervals in a single flagellar motor switch are best described by gamma distributions, which indicate the the presence of multiple hidden Markov steps within a motor, even when decoupled from its singalling network \cite{bibKorobkovaPRL}. For the sake of illustrating the utility of our formalism, let us consider gamma-distributed runs and estimate the drift velocity of the bacterium. Let $\xi_n(t)$ be the probability distribution of run intervals (assume tumbles 
to be instantaneous), which we take to have the form 

\begin{eqnarray} \label{fT_Gamma}
 \xi_n(t) &=& \frac{R^n}{\Gamma(n)}t^{n-1}e^{-Rt} H(t) \nonumber \\
      &=& (-1)^{n-1}\frac{R^{n-1}}{\Gamma(n)}\frac{d^{n-1}}{dR^{n-1}}\Big(\xi_1(t) \Big)
\end{eqnarray}
where $\xi_1(t)=Re^{-Rt} H(t)$, $n\geq 1$ is the number of hidden steps in a single $ CW \leftrightarrow CCW $ switch. 
The corresponding cumulative probability  is given by $f_n(t)=\int_0^t\xi_n(T)dT$, whose 
Fourier transform, from Eq.\ref{fT_Gamma}, is given by 

\begin{equation} \label{Fw_Gamma}
 F_n(\omega) = \hat{\mathcal{F}}_{R}^nF_1(\omega)
 \label{eqFomega}
\end{equation}

for $n>1$,  where 

\begin{equation}
\hat{\mathcal{F}}_{R}^n=(-1)^{n-1}\frac{R^{n-1}}{ \Gamma(n)}\frac{d^{n-1}}{dR^{n-1}}
\label{eqoperator}
\end{equation}

is a linear differential operator, $\Gamma(n)=(n-1)!$ being the standard gamma function. From the linearity of the relation in Eq.\ref{Fw_Gamma}, it follows that if the integrals in Eq.\ref{eq5b} are replaced with $I^{(n)}_{1/2}(s;F_n)= \hat{\mathcal{F}}_{R}^n I_{1/2}(s;F_1)$, the l.h.s. of the equation becomes the new diffusion coefficient, to be denoted $D^{(n)}$, which gives 

\begin{equation}
v_1^{(n)}\sim 2\alpha\kappa R\Delta D^{(n)}
\label{eqX1A}
\end{equation}

as the generalisation of Eq. \ref{eq12+}. It may be shown easily that 

\begin{equation}
I_1^{(n)}(0;F)=1/R~~;~~I_2^{(n)}=n/R^2,
\label{eqI1I2}
\end{equation}
and hence $D^{(n)}=nD^{(1)}$, where $D^{(1)}=D_e$, the latter as given in Eq.\ref{eq6+}. Next, we use the more general expression for $v_2$ in Eq.\ref{eq19} (Sect.~\ref{sec:5d}), and use Eq.\ref{eqFomega}, leading to 

\begin{equation} \label{v2n}
 v_2^{(n)}\sim \frac{R\kappa \alpha v^2}{2I_1^{(n)}(0;F)}  \hat{\mathcal{F}}_{R}^{n} 
 \big[(1-2\Delta R - e^{-\Delta R})/R^3\big].
 \label{eqX2A}
\end{equation}

which is the required generalisation of Eq. \ref{eq13} for arbitrary $n$. 

\begin{figure}[!h]
  \begin{center}
  {
  \includegraphics[scale=0.6]{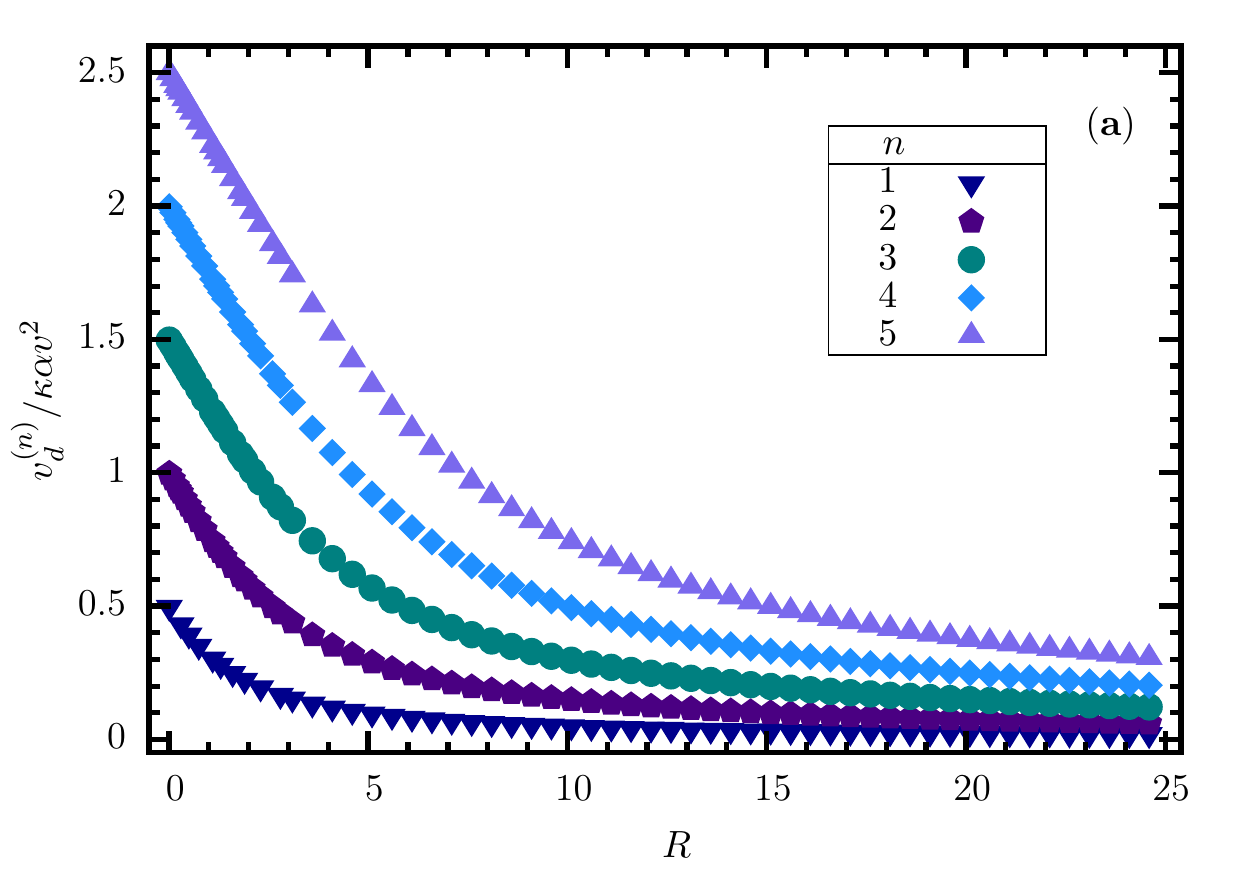}
  \includegraphics[scale=0.6]{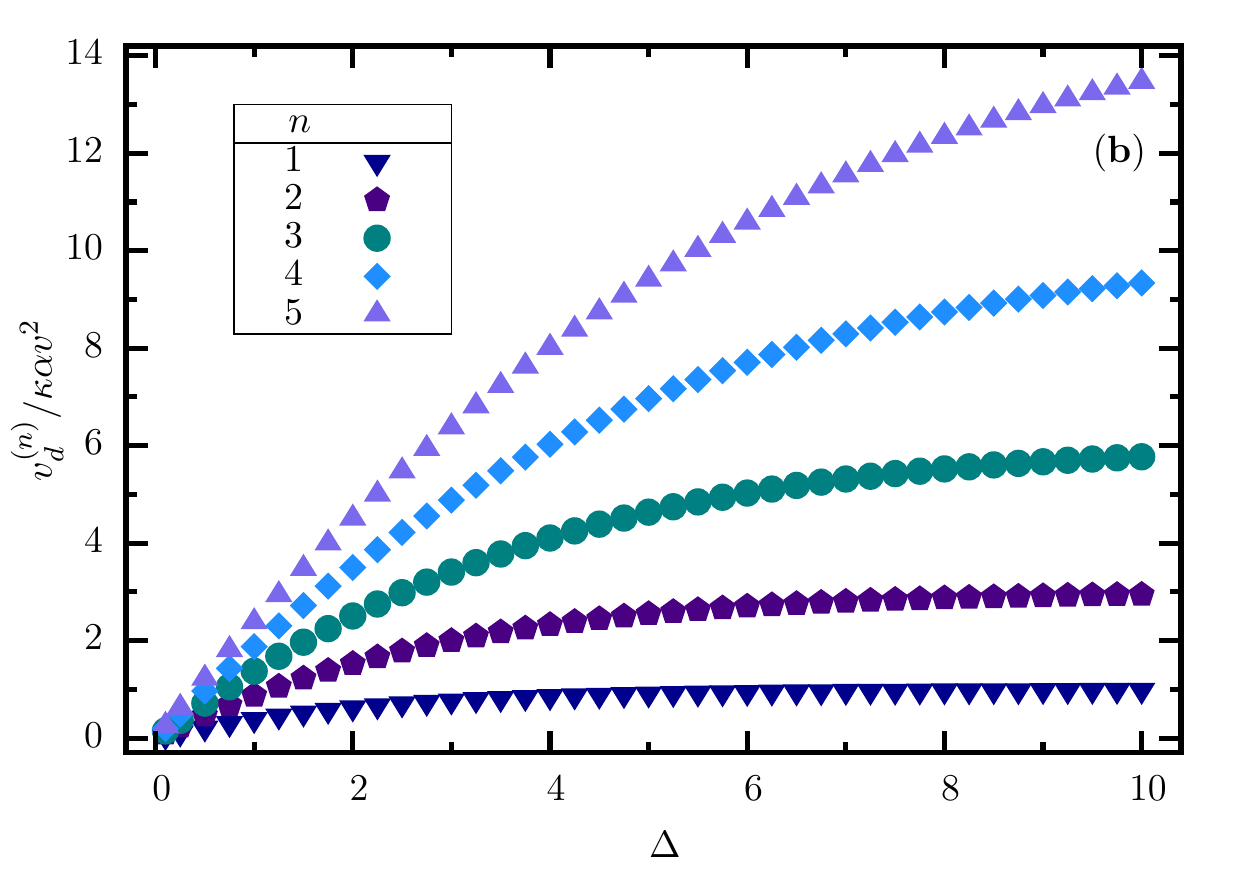}
  }
  \caption{The scaled drift velocity $v_d^{(n)} / \kappa \alpha v^2$ for gamma-distributed run-lengths and instantaneous tumbles as a 
  function of (a) the Poisson rate $R$ (in s$^{-1}$) and (b) the time delay $\Delta$ between the positive and negative lobes of the response
  function. In (a), $\Delta=1.0s$ and in (b),  $R=0.5$ s$^{-1}$. }
  \label{Plot1}
  \end{center}
  \end{figure}

The net drift velocity $v_d^{(n)}$ is given by the sum of the expressions in Eq.\ref{eqX1A} and Eq.\ref{eqX2A}. 
In Fig.~\ref{Plot1}, the scaled drift velocity $v_d^{(n)}/\kappa \alpha v^2$ computed using Eq.\ref{eqX1A} and Eq.\ref{eqX2A} is plotted as a function of the Poisson rate $R$ (a) and time delay $\Delta$ (b),  for various integers $n$. We find that the drift velocity is an increasing function of $n$, but the qualitative nature of the variation with $R$ or $\Delta$ remains the same for all $n$.

\section{Summary and conclusions}

In this paper, we have presented a path-integral method to compute various dynamical quantities of interest in run and tumble motion of 
bacteria, with or without chemotaxis. Similar to a few early papers\cite{bibReneaux,bibCelani,bibSakuntala},  our study is also motivated by the pioneering work of de Gennes \cite{bibdeGennes}. We show that de Gennes' elegant result for the 
drift velocity in a weak linear gradient can be reproduced as the first term in a systematic perturbative expansion in powers of the attractant gradient, 
by computing the mean displacement over a single trajectory (rather than a single run), and then averaging over all trajectories. The formalism also naturally includes directional correlations between runs; here, we predict that, for unbiased motion, positive correlation between directions of 
successive runs increases the effective diffusion coefficient, while negative correlation reduces it.  Most importantly, in its 
general form, the method can handle non-exponentially distributed run and tumble durations (likely relevant for {\it E.coli}, as indicated by experimental data). Although this may not be directly relevant for bacterial run and tumble motion, we have also shown that the formalism predicts correctly 
the occurrence of ballistic, super-diffusive and diffusive behaviour of the mean square displacement when the run intervals are algebraically 
distributed. 

As an illustration of the utility of our formalism, we have computed the drift velocity and diffusion coefficient of chemotaxing bacteria when the run interval durations are gamma-distributed (the tumbles treated as instantaneous 
events). Such gamma distributions arise naturally when there are hidden steps in the run-tumble transition; 
the total number of such steps may be denoted by $n$. For $n>1$, the distribution has a maximum, which becomes sharper with increase in $n$. 
We find that the drift velocity is a monotonically increasing function of $n$, and a decreasing function of the tumble rate for all $n$. This investigation was motivated by some 
recent experiments, where  the distributions of clockwise and counter-clockwise intervals of a single flagellar motor in an 
immobilized bacterium were measured as a function of the mean clockwise bias of the motor. It was found that both intervals are, in general, 
gamma-distributed, but the number of hidden steps in each transition (CW$\to$ CCW and vice-versa) depends continuously on the bias\cite{bibKorobkovaPRL,bibParkBiophysJ}. 
Although the CCW and CW spinning states  of a single motor may not directly correspond to run and tumble events of a bacterium, it is likely that the latter also displays similar statistical 
behaviour. We hope that a future experiment may explore the statistics of run and tumble durations in more detail. Likewise, the model we used here assumes that run durations are modulated by an attractant gradient, but not tumbles. This is consistent with the prevailing picture of chemotaxis in {\it E. coli}, but if a future experiment were to indicate otherwise, the formalism is equipped to handle it as well. 

Even though we have not been able to extend the computation of drift velocity (or diffusion coefficient) to higher orders in $\nabla c$ so far because of  computational complexity, this should certainly be possible and would be one of our goals for the immediate future. In the present paper, we have also limited our attention to simple mean quantities describing the cell's motion, like drift velocity and diffusion coefficient, but many others, e.g., probability distribution of the number of tumble events, with and without chemotaxis, can be calculated, in principle. Other quantities of general interest, which could be computed from our model, include the correlation between successive run (and tumble) durations in a chemotaxing cell. 

\begin{acknowledgements}
M.G would like to thank R. Adhikari for helpful discussions in the early stages of this work. 
\end{acknowledgements}


%
%



%
%
\bibliographystyle{spphys}
\bibliography{RunTumble}

%
%
%
%



\section{Appendix}

\subsection{Probability distribution of displacement in the exponential model}
\label{sec:5a}

Here, we calculate the probability distribution of the displacement vector ${\bf r}$ of the bacterium after time $t$. Note that 
for a given trajectory with $N$ tumbles in all, the displacement vector is given by the expression in Eq.\ref{eqRN}. 
The probability distribution $P({\bf r},t)$ is evaluated using the functional in Eq.\ref{eq6a}; 
$P({\bf r},t)=\langle \delta^{(2)}({\bf r}-{\bf R}_N({\bf T}, {\bm \theta}))\rangle_e$, whose Fourier transform $P({\bf k},t)=(2\pi)^{-1}\int P({\bf r},t)e^{i{\bf k}\cdot{\bf r}}$ turns out to be

%
%
%
%
%

\begin{eqnarray}
P(\vec{k},t)=\sum_{N=0}^{\infty}\frac{1}{(2\pi)^{N+2}} R^N e^{-At} \int_{0}^{2\pi} d\theta_1...d\theta_{N+1} \int_{0}^{t} dT_1 e^{i\alpha_1 T_1}\int_{0}^{t-T_1} dT_2 e^{i\alpha_2 T_2}....\nonumber\\
\int_{0}^{t-\sum_{j=1}^{N-1}T_j} dT_N e^{i\alpha_N T_N}~~~
\label{eq5a1}
\end{eqnarray}
where $A = R+ i v\vec{k}\cdot {\bf e}_{N+1}$ and $\alpha_j = v\vec{k}\cdot({\bf e}_{N+1}-{\bf e}_{j})$. After using the generalised convolution theorem in Sect.~\ref{sec:1}, we find the Laplace-transformed distribution $P({\bf k},s)=\int_{0}^{\infty} P({\bf k},t)e^{-st}dt$:

%
%
\begin{equation}
\tilde{P}(\vec{k},s)=\sum_{N=0}^{\infty}\frac{1}{(2\pi)^{N+2}} R^N  \left[\int_{0}^{2\pi} \frac{d\theta}{s+R+i v \vec{k}\cdot {\bf e}(\theta)}\right]^{N+1}
\label{eq5a2}
\end{equation}

where ${\bf e}(\theta)=\cos\theta {\bf i}+\sin\theta {\bf j}$ are unit vectors. After carrying out the straightforward angular integration, we find 

\begin{equation}\label{eqA1}
\tilde{P}({\bf k},s)=\frac{1}{2\pi}\Bigg( \frac{1}{\sqrt{(s+R)^2+v^2k^2} - R}\Bigg),
\label{eq5a3}
\end{equation}
where $k=|{\bf k}|$. For small $k$, one can expand the denominator in powers of $k^2$. After carrying out the $s\to t$ inverse 
Laplace transform of the resulting expression, we find  
%
%
%

\begin{equation}
\tilde{P}({\bf k},t)\sim \frac{1}{2 \pi}\exp \bigg({\frac{-v^2 k^2t}{2R}}\bigg).~~~~(vk/R\to 0)
\label{eq5a4}
\end{equation}

The inverse Fourier transform (${\bf k} \longrightarrow {\bf r}$) of Eq.\ref{eq5a4} yields the large distance asymptotic form 

\begin{equation}
P({\bf r},t)\sim \frac{1}{ 4 \pi D_e t} \exp\bigg(\frac{-r^2}{4D_et}\bigg), ~~~~(r\gg v/R)
\end{equation}

where $r=|{\bf r}|$ is the net displacement and the diffusion coefficient $D_e$ is given in Eq.\ref{eq6+}. 

\subsection{Mean square displacement for L\'evy-like (algebraic) $f(T)$}
\label{sec:5c}


{\it Case I: $0<\beta\leq 1$}\\

The Fourier transform of the function $f(T)=(1+\gamma T)^{-\beta}$ is defined as
\begin{equation}
 F(\omega) = \frac{1}{2\pi} \int_0^\infty \frac{e^{i\omega T}}{(1+\gamma T)^{\beta}}dT
 \label{eq5c1}
\end{equation}

Substituting $1+\gamma T = x$ in the above integral, we obtain

\begin{equation}
 F(\omega)= \frac{e^{-i \omega/\gamma}}{2\pi \gamma}\int_1^\infty \frac{e^{i \omega x/\gamma}}{x^\beta}dx=F^{\prime}(\omega)-F^{\prime\prime}(\omega), 
 \label{eq5c2}
\end{equation}

where 

\begin{equation}
F^{\prime}(\omega)= \frac{e^{-i \omega/\gamma}}{2\pi \gamma}\int_0^\infty \frac{e^{i \omega x/\gamma}}{x^\beta}dx~;~~ F^{\prime\prime}(\omega)=\frac{e^{-i \omega/\gamma}}{2\pi \gamma}\int_0^1 \frac{e^{i \omega x/\gamma}}{x^\beta}dx
\label{eq5c2+}
\end{equation}

\begin{figure}
\begin{subfigure}{.5\textwidth}
  \centering
  \includegraphics[width=1.5\linewidth]{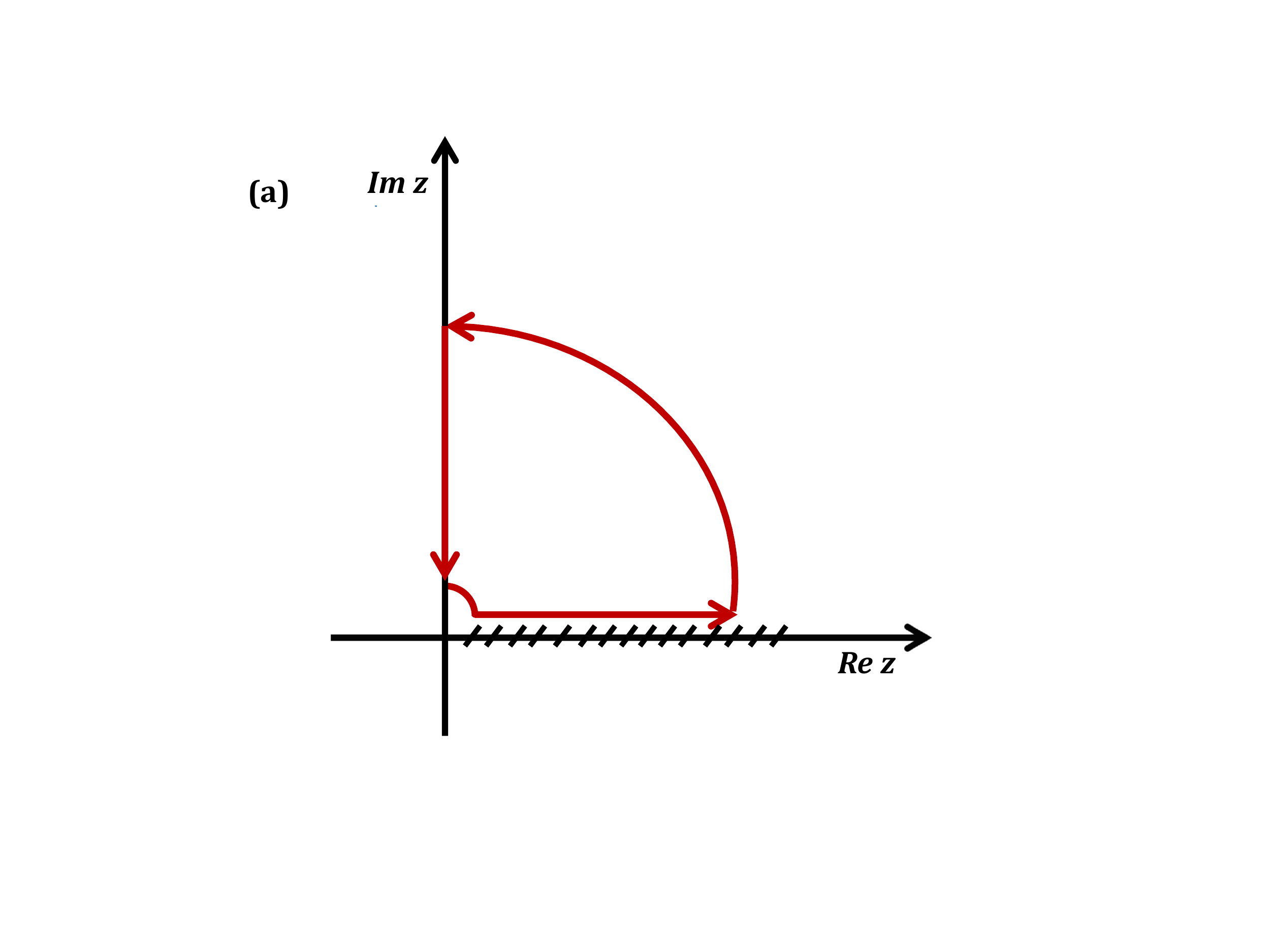}
  \label{fig:sfig1}
\end{subfigure}%
\hspace{-0.5cm}
\begin{subfigure}{.5\textwidth}
  \centering
  \includegraphics[width=1.5\linewidth]{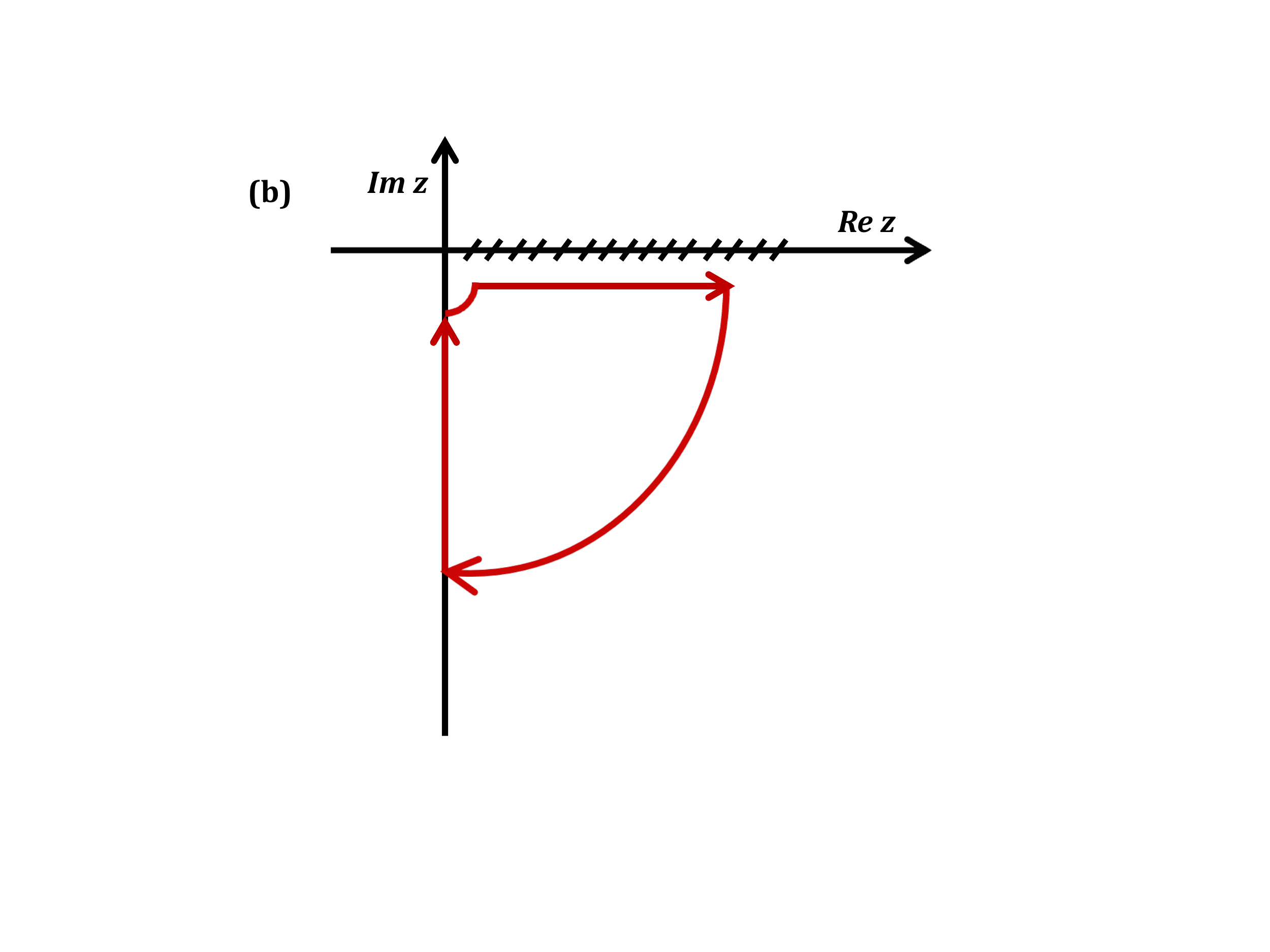}
  \label{fig:sfig2}
\end{subfigure}
\vspace{-1.5cm}
\caption{The contours used for evaluating the integrals in (a) Eq.(\ref{Fz}) and (b) Eq.(\ref{IA}).}
\label{fig:fig}
\end{figure}

Note that $F^{\prime\prime}(\omega)$ does not diverge as $\omega\to 0$, while $F^{\prime}(\omega)$ does (see below); 
therefore in the long-time limit we are interested in, $F(\omega)\sim F^{\prime}(\omega)$ . Consider now the complex-valued integral 

\begin{equation} \label{Fz}
{\mathcal F}(z)=\oint \frac{e^{i\omega z}}{z^{\beta}}dz
\label{eq5c3}
\end{equation}
with the contour of evaluation chosen as in Fig.\ref{fig:fig}(a), where $z=x+iy$. In the limit $R\to\infty$, the integral over 
the quarter-circle vanishes according to Jordan's lemma, and we arrive at 


\begin{equation}
 \int_0^\infty \frac{e^{i \omega x/\gamma}}{x^\gamma}dx + \int_\infty^0 \frac{e^{i \omega (iy)/\gamma}}{(iy)^\gamma}idy = 0
\label{eq5c4} 
\end{equation}
where the r.h.s. is zero since ${\mathcal F}(z)$ has no pole inside the contour. Therefore

\begin{equation} \label{Fw}
 F(\omega)\sim \frac{e^{-i\omega/\gamma}}{2\pi \gamma^\beta}i^{1-\beta}\omega^{\beta-1}\Gamma(1-\beta)
\label{eq5c5} 
\end{equation}

The integral $I_1(s;F)$ in Eq.(\ref{I1I3}) becomes 

\begin{equation}
 I_1(s;F)=\frac{\Gamma(1-\beta)\gamma^{-\beta}e^{\frac{i\pi}{2}(1-\beta)}}{\pi}I_A(s)
 \label{eq5c6}
\end{equation}
where 

\begin{equation} \label{IA}
 I_A(s) = \int_{0}^{\infty} \frac{e^{-i\omega/\gamma}}{s+i\omega}\omega^{\beta-1}d\omega.
 \label{eq5c7}
 \end{equation}

To evaluate the integral, we use the contour in Fig.\ref{fig:fig}(b); along the branch cut, $\omega$ will be replaced by 
$\omega e^{i2\pi}$, while along the imaginary axis, $\omega=iy$. Applying Cauchy's residue theorem again, we find 
%

 \begin{equation}
e^{i2\pi(\beta-1)}I_A + i^\beta \int_{-\infty}^0 y^{\beta-1} \frac{e^{y/\gamma}}{s-y}dy = 0
\label{eq5c8}
\end{equation}

Substituting $\xi=-y$, we find 

\begin{equation}
 I_A = e^{-i\pi\beta/2} \int_0^\infty \xi^{\beta-1} \frac{e^{-\xi/\gamma}}{s+\xi}d\xi
\label{eq5c9} 
\end{equation}

The integral in the r.h.s of Eq.\ref{eq5c9} can be evaluated by introducing an auxiliary variable $\lambda$ through the integral representation $1/(s+\xi) = \int_0^\infty e^{-\lambda(s+\xi)}d\lambda $. After substituting in the above equation, we find 


\begin{equation}
\int_0^\infty \xi^{\beta-1} \frac{e^{-\xi/\gamma}}{s+\xi}d\xi= \Gamma(\beta) \gamma^{\beta} s^{\beta-1}\int_0^\infty \frac{e^{-\eta}}{(s+\gamma \eta)^\beta}d\eta
\label{eq5c10}
\end{equation}
where $\eta=\lambda s$. In the limit $s\to 0$, the integral becomes $\Gamma(1-\beta)$. We now substitute the resulting limiting expression 
for $I_A(s)$ 
in Eq.\ref{eq5c6} to find that $I_1(s;F)\propto s^{\beta-1}$ as $s\to 0$. It follows that $I_2(s;F)\propto s^{\beta-2}$ and hence, 
from Eq.\ref{eq5b}, we find 
$\langle r^2\rangle\propto t^2$ for large $t$. The motion is, therefore, ballistic in this regime. \\

%

{\it Case II: $1< \beta<2$}\\

For $\beta>1$, the Fourier transform of the survival probability can be expressed as

\begin{equation}
 F_{\beta}(\omega)=\frac{1}{2\pi\gamma(1-\beta)}\int_0^\infty e^{i\omega t} d[(1+\gamma t)^{1-\beta}]
 \label{eq5c11}
\end{equation}
which can be simplified to

\begin{equation}
 F_\beta(\omega) = \frac{1}{2\pi\gamma(\beta-1)} \Big[ 1+i\omega F_{\beta -1}(\omega) \Big]. 
\label{eq5c12}
\end{equation}

For $\beta<2$, after using the expression given in (\ref{eq5c5}), we get

\begin{equation}
 F_\beta(\omega) = \frac{1}{2\pi\gamma(\beta-1)}\Big[ 1+\Gamma(2-\beta)e^{i\pi(3-\beta)/2}e^{-i\omega/\gamma}(\omega/\gamma)^{\beta-1}\Big]
 \label{eq5c13}
\end{equation}

For this case, one can write $I_1(s;F)=I_1^{(1)}+I_1^{(2)}$, which are defined as follows: 

\begin{eqnarray}
 I_1^{(1)} &=& \int_{-\infty}^{\infty} \frac{d\omega}{2\pi\gamma(\beta-1)(s+i\omega)} = \frac{-i}{2\pi\gamma(\beta-1)}\equiv A_1\nonumber \\
 I_1^{(2)} &=& \frac{\Gamma(2-\beta)e^{i\pi(3-\beta)/2}}{2\pi \gamma^\beta (\beta-1)}
 \int_{-\infty}^{\infty} \frac{e^{-i\omega/\gamma}\omega^{\beta-1}}{s+i\omega}d\omega= A_2 s^{\beta - 1}
\end{eqnarray}
where $A_1$ and $A_2$ are constants. Therefore $I_1(s;F)= A_1 + A_2s^{\beta-1} $ and $I_2\propto s^{\beta - 2}$. 
From Eq.\ref{eq5b}, it follows that ${\mathcal L}_s[\langle r^2 \rangle] \propto s^{4-\beta}$, and hence 
$\langle r^2\rangle \propto t^{3-\beta}$.
Thus, the run and tumble motion is super-diffusive in this regime. \\

{\it Case III: $\beta\geq 2$}\\

Here, we apply the recursion relation in Eq.\ref{eq5c12} one more time to find that $I_1(s;F)=B_1+B_2 s+B_3 s^{\beta-1}$,
where $B_1,B_2,B_3$ are non-zero constants. It then follows that in the long time limit, $\langle r^2\rangle\propto t$,  i.e., 
the motion is purely diffusive in this regime.

\subsection{Mean square displacement in unbiased motion with directional persistence}
\label{sec:5b}


The Fourier-Laplace transform of the probability distribution $P({\bf r},t)$ of the position ${\bf r}$ at time $t$ is given by the 
following  generalisation of Eq.\ref{eq5a2}: 

\begin{equation}
\tilde{P}({\bf k},s)= \sum_{N=0}^{\infty}R^N \int {\mathcal D}\theta \frac{\psi(\theta_1)}{s+R+ i v {\bf k}\cdot{\bf e}_1}\prod_{j=1}^{{N+1}}\frac{G(\theta_{j+1}|\theta_j)}{s+R+ i v {\bf k}\cdot {\bf e}_j}, 
\label{eq5b3}
\end{equation}

where, the unit vectors ${\bf e}_j$ have been defined following Eq.\ref{eqRN}. Let us now define a set of $N$ integrals, 


\begin{eqnarray}
I_{1}(\theta_N)  &=& \int_{0}^{2\pi}\frac{d\theta_{N+1}G(\theta_{N+1}\mid\theta_{N})}{s+R+ i v {\bf k}\cdot{\bf e}_{N+1}}\nonumber\\
I_2(\theta_{N-1})  &=& \int_{0}^{2\pi}\frac{d\theta_{N}G(\theta_{N}\mid\theta_{N-1})I_1(\theta_N)}{s+R+ i v {\bf k}\cdot{\bf e}_{N}}\nonumber\\
&\vdots& \nonumber\\
I_N(\theta_1)  &=& \int_{0}^{2\pi}\frac{d\theta_{2}G(\theta_{2}\mid\theta_{1})I_{N-1}(\theta_2)}{s+R+ i v {\bf k}\cdot{\bf e}_{2}}.\nonumber\\
\label{eq5b4}
\end{eqnarray}

After substituting Eq.\ref{eq5b2} in Eq.\ref{eq5b4}, it turns out that, for general $n$, the integral $I_n$ can be expressed as 

\begin{equation}
I_n = C_n + D_n\cos(\theta_{N-n+1}), 
\label{eq5b5}
\end{equation}

where 


\begin{equation}
\left[ \begin{array}{c} C_n \\ D_n  \end{array} \right]=
\left[ \begin{array}{cc} (s+R)A_1 & -ivkA_2  \\
-ivkJA_2 & (s+R)JA_2 \end{array} \right]^{n-1}
\left[\begin{array}{c} C_1 \\ D_1 \end{array} \right];\quad \forall n \geq 2
\label{eq5b6}
\end{equation}

and $k\equiv |{\bf k}|$. The constants $A_1$ and $A_2$ in the above equation are given by 

\begin{eqnarray}
A_1 &=& \frac{1}{2\pi}\int_{0}^{2\pi}\frac{d\theta}{(s+R)^2 + v^2k^2\cos^2\theta}=\frac{1}{(s+R)\sqrt{(s+R)^2 +v^2k^2}}\nonumber\\
A_2 &=& \frac{1}{2\pi}\int_{0}^{2\pi}\frac{\cos^2\theta d\theta}{(s+R)^2 + v^2k^2\cos^2\theta}=\frac{1}{v^2k^2}\bigg[1-\frac{s+R}{\sqrt{(s+R)^2 +v^2k^2}}\bigg]. 
\label{eq5b7}
\end{eqnarray}
where $C_1=(s+R)A_1$ and $D_1 = - i v k J A_2$. The r.h.s of Eq.\ref{eq5b6} is evaluated using Cayley-Hamilton theorem, which uses the following eigenvalues of the $2\times 2$ 
matrix in Eq.\ref{eq5b6}:

\begin{eqnarray}
\lambda_1 = \frac{(s+R)(A_1+JA_2)+\sqrt{(s+R)^2(A_1+JA_2)^2-4J[(s+R)^2A_1A_2+v^2k^2A_2^2]}}{2}\nonumber\\
\lambda_2= \frac{(s+R)(A_1+JA_2)-\sqrt{(s+R)^2(A_1+JA_2)^2-4J[(s+R)^2A_1A_2+v^2k^2A_2^2]}}{2}
\label{eq5b8}
\end{eqnarray}

The final exact result, after some simplifications, is 

\begin{eqnarray}
\tilde{P}({\bf k},s)
=\frac{1}{2\pi}\bigg(\frac{1}{\sqrt{(s+R)^2+v^2k^2}-R}\bigg)
\nonumber\\
-\frac{JRv^2k^2 A_2^2}{2\pi\bigg[(s+R)^2A_1^2 - J \bigg( 2 v^2 k^2 A_2^2 + (s+R)^2 A_1 A_2 \bigg)\bigg]\bigg(\sqrt{(s+R)^2+v^2k^2}-R\bigg)^2}
\label{eq5b9}
\end{eqnarray}

which, as expected, reduces to Eq.\ref{eq5a3} when $J=0$. The Laplace transform of the mean square displacement 
$\langle r^2\rangle=\langle x^2\rangle+\langle y^2\rangle$ is given by 

\begin{equation}
{\mathcal{L}}_s[\langle r^2\rangle]=2\pi\bigg\{\frac{\partial^2\tilde{P}({\bf k},s)}{\partial (-ik_x)^2}\bigg\vert_{{\bf k}=0} +\frac{\partial^2\tilde{P}({\bf k},s)}{\partial (-ik_y)^2}\bigg\vert_{{\bf k}=0}\bigg\},
\label{eq5b10}
\end{equation}

which we use in Eq.\ref{eq5b9} to find that ${\mathcal{L}}_s[\langle r^2\rangle]\sim 4D_J/s^2$ as $s\to 0$, or, 
equivalently, $\langle r^2\rangle\sim 4D_Jt$ as $t\to\infty$, with the diffusion coefficient $D_J$ given in Eq.\ref{eq5b11}. 

\subsection{Diffusion coefficient and drift velocity for general $f(T)$} 
\label{sec:5d}

In this section, we present the calculation of drift velocity under chemotaxis,  for arbitrary run interval distribution $f(T)$, but assuming that the unbiased motion is diffusive in the long-time limit. Let us start from Eq.\ref{eq5}: the attractant gradient modifies the survival probability in the run state, which 
we express in the general form

\begin{equation}
f(T_i; {\bf T}, {\bm\theta})=f^{(0)}(T_i)\left\{1-\int_{t_{i-1}}^{t_i}\delta R (t;{\bf T},{\bm \theta})dt+.....\right\}
\label{eq15}
\end{equation}
where 

\begin{equation}
 \delta R (t;{\bf T},{\bm \theta}) =-\alpha \overline{R}\int_{0}^{t}\chi(t-t^{\prime})x(t^{\prime})dt^{\prime}
\label{eq15+}
\end{equation}

is the perturbation due to the gradient and $\overline{R}$ is a baseline switch rate. From Eq.\ref{eq15}, we find 

\begin{equation}
{\dot f}(T_i; {\bf T},{\bm\theta})={\dot f^{(0)}}(T_i)-\delta f_i  ({\bf T},{\bm \theta})
\label{eq15a}
\end{equation}
where the dot denotes differentiation with respect to the first variable, and 

\begin{equation}
\delta f_i  ({\bf T},{\bm \theta}) ={\dot f^{(0)}}(T_i)\int_{t_{i-1}}^{t_i}\delta R (t;{\bf T},{\bm \theta}) dt+f^{(0)}(T_i)\delta R (t_i;{\bf T},{\bm \theta})
\label{eq15b}
\end{equation}

After using Eq.\ref{eq15} and Eq.\ref{eq15b}, the probability functional in Eq.\ref{eq5} may be expanded in the form

\begin{eqnarray}
\Phi_{N}^{(m,c)}({\bf T},{\bm\theta};t)=\frac{1}{(2\pi)^{N+1}}f^{(0)}(T_{N+1})\delta\left(T_{N+1}+\sum_{i=1}^{N}T_i-t \right)\prod_{i=1}^{N}(-1)^N{\dot f^{(0)}}(T_i)\bigg[1-\nonumber\\
\int_{0}^{t}\delta R (t^{\prime};{\bf T},{\bm \theta})dt^{\prime}-\sum_{i=1}^{N}\delta R (t_i;{\bf T},{\bm \theta})\frac{f^{(0)}(T_i)}{{\dot f^{(0)}}(T_i)}+..........\bigg],~~
\label{eq16}
\end{eqnarray}
which replaces Eq.\ref{eq11}, for general $f(T)$. The mean position $\langle x(t)\rangle=x_1(t)+x_2(t)$ again, with slightly modified 
expressions: 

\begin{eqnarray}
x_1(t) &=& \alpha\kappa \overline{R}\int_{t-\Delta}^{t}\langle x(t^{\prime})x(t)\rangle dt^{\prime}\nonumber\\
x_2(t) &=& -\alpha\kappa\overline{R}\sum_{j=1}^{N}\bigg\langle x(t)[x(t_j)-x(t_j-\Delta)]\frac{f^{(0)}(T_j)}{{\dot f^{(0)}}(T_j)}\bigg\rangle
\label{eq17}
\end{eqnarray}
where the averages are to be computed using the distribution function in Eq.\ref{eq5}. Note that for $f(T)=e^{-RT}H(T)$, 
the expressions in Eq.\ref{eq17} reduce to those in Eq.\ref{eq12a} and Eq.\ref{eq12}. We again use the standard result 
$\langle x(t)x(t^{\prime})\rangle\sim 2Dt^{\prime}$ (for $t^{\prime}\leq t$) in the long time limit to find $x_1(t)\sim v_1 t$ where 

\begin{equation}
v_1 = 2\kappa {\overline R}\alpha D\Delta 
\label{eq17+}
\end{equation}

with $D$ given by Eq.\ref{eq6}. The Laplace transform of $x_2(t)$ turns out to be 

\begin{equation}
{\mathcal L}_s[x_2]=-\frac{\kappa\overline{R}\alpha v^2}{2}\sum_{N=0}^{\infty}\int {\mathcal D}{\omega}\sum_{i=1}^{N}(N-i+1)\beta_i(s+i\omega_i)F(\omega_i)\prod_{j\neq i}i\omega_jF(\omega_j), 
\label{eq18}
\end{equation}

which is a generalisation of Eq.\ref{eq5d12} in Sect.~\ref{sec:5d}. The explicit expressions for the integrals $\beta_i(s)$ are given 
in Eq.\ref{eq5d10} and Eq.\ref{eq5d11}. After completing the summation, we find that $x_2(t)\sim v_2 t$ for large $t$, where

\begin{equation} \label{v2_Fw}
v_2=-\frac{{\overline R}\kappa\alpha v^2}{2 I_1(0;F)}\int d\omega F(\omega)\bigg\{-\frac{2\Delta}{\omega^2}+\frac{1-e^{-i\omega\Delta}}{i \omega^3}\bigg\}
\label{eq19}
\end{equation}

It may be easily verified that, for the exponential model with $F(\omega)=[2\pi(R-i\omega)]^{-1}$ (and the substitution $\overline{R}\to R$), the r.h.s of Eq.\ref{eq19} reduces to that of Eq.\ref{eq13}, as expected. The net drift velocity is given by the sum of the expressions in Eq.\ref{eq17+} and Eq.\ref{eq19}:

\begin{equation}
v_d= \frac{{\overline R}\kappa\alpha v^2}{2 I_1(0;F)}\int d\omega F(\omega)\frac{(e^{-i\omega\Delta}-1)}{i\omega^3}
\label{eq19a}
\end{equation}

which generalises Eq.\ref{eq14} for arbitrary $f(T)$.

\end{document}